\documentstyle[11pt,aaspp4]{article}

\begin{document}
\def\jh{\hbox{$^{\rm h}\!$}}
\def\jm{\hbox{$^{\rm m}\!$}}
\def\js{\hbox{$.\!\!^{\rm s}$}}
\def\pjh{\phantom{\hbox{$^{\rm h}\!$}}}
\def\pjm{\phantom{\hbox{$^{\rm m}\!$}}}
\def\pjs{\hbox{$.\!\!^{\phantom{\rm s}}$}}
\def\jd{\hbox{$\!^\circ$}}
\def\pjd{\phantom{\hbox{$\!^\circ$}}}
\def\jam{\hbox{$^\prime$}}
\def\pjam{\phantom{\hbox{$^\prime$}}}
\def\jas{\hbox{\kern 0.13ex.\kern -0.95ex%
  \raisebox{.9ex}{\scriptsize$\prime\prime$}\kern -0.1ex}}
\def\pjas{\hbox{\kern 0.13ex.\kern -0.95ex%
  \raisebox{.9ex}{\phantom{\scriptsize$\prime\prime$}}\kern -0.1ex}}
\def\ppc{\phantom{$\%$}}

\title {A DEEP MULTICOLOR SURVEY. III. ADDITIONAL SPECTROSCOPY AND
IMPLICATIONS FOR THE NUMBER COUNTS OF FAINT QUASARS}

\author {Julia D. Kennefick and Patrick S. Osmer}
\affil{Astronomy Department,
The Ohio State University, 174 West 18th Avenue, Columbus, OH 43210;
julia@astronomy.ohio-state.edu and posmer@astronomy.ohio-state.edu}

\author {Patrick B. Hall}
\affil{Steward Observatory, University of Arizona, AZ 85721;
phall@as.arizona.edu}

\author {Richard F. Green}
\affil{National Optical Astronomy
Observatories, P.O. Box 26732, Tucson, AZ 85726; rgreen@noao.edu}

\begin{abstract}

We have made spectroscopic identifications of 39 additional quasar 
candidates from the Deep Multicolor Survey (DMS)
of Hall {\it et al.} (1996, ApJ, 462, 614).  
We have identified 9 new quasars with $0.3 < z < 2.8$ 
and $16.8 < B < 21.6$, all from the group of candidates with 
ultraviolet excess (UVX).  No new quasars with $z > 3$ were 
found among the observed candidates selected due to their red
($B-R$) and ($V-R$) colors.  As a 
result, there are now 55 confirmed quasars in the survey: 42 
with $0.3 < z < 2$, nine with $2 < z < 3$, three with $3 < z < 4$, 
and 1 at $z = 4.3$.  One new quasar, DMS 0059$-$0055, is very
bright with $B=16.8$ and $z=0.3$, making its detection by
our survey very unexpected.  Including this new spectroscopy, the results
of the DMS are converging with the predicted space densities
of other surveys.  In particular, we no longer find an excess of 
quasars with $z<2.3$ and $B<21$ in the survey
over predictions based on models by Koo \& Kron.  Also, the excess
in the number of quasars seen at $z>3$ over predictions based
on models by Warren, Hewett,
\& Osmer is less than previously suggested.
We also demonstrate the success of our quasar color modeling which is
important in assessing the completeness of our survey.

\end{abstract}

\newpage

\section{Introduction}

This is the third in a series of papers describing results
from the Deep Multicolor CCD Survey (DMS) of \cite{HALL2} (1996ab;
Papers I and II) conducted at the KPNO 4m telescope.  
The survey covered 0.83 deg$^2$ using six filters
(U,B,V,R,I75,I86) covering the range from 0.34 to 0.86$\mu$m
in six fields at high galactic latitude.  The imaging
data have average 5$\sigma$ limiting detection magnitudes 
ranging from 22.1 to 23.8. 
The motivation of the survey is to search for lower
luminosity quasars at $z > 3$ than have previously been studied and to
search for lower luminosity quasars with $z \approx 2$ to constrain the
nature
of the evolution of the luminosity function.  In addition, the survey
is valuable for the study of faint field galaxies and stars in the
galactic halo ({\it e.g.}, \cite{LIU}.)

Paper I contains the details of the construction of the stellar catalog
of 21375 objects, including
a detailed description of the imaging observations and reductions, object
classification, photometry, astrometry, and the catalog completeness and
contamination.  
Paper II describes the search for quasars including
candidate selection criteria, survey completeness estimates,
and initial spectroscopy which resulted in the discovery
of 46 quasars, including one at $z=4.3$, and a comparable
number of compact narrow emission line galaxies (CNELG's).

The initial spectroscopy described in Paper II was used along with determinations
of the survey efficiency to estimate the number of quasars contained in the survey
for different redshift and magnitude bins.  Our expected numbers were compared
with predictions based on models from \cite{KK88} (1988, hereafter KK88) and 
\cite{WHO} (1994, hereafter WHO).  
It was concluded that the DMS contains more quasars with $B < 21$
and $z<2.3$ than expected from the results of Koo \& Kron and more quasars
at $z>3$ than expected from the results of Warren, Hewett, \& Osmer.
However, the initial spectroscopy included less than half of the candidates at the
brightest magnitudes and even fewer of the fainter candidates.

Here we report additional follow-up
spectroscopy obtained at the KPNO 4m telescope in April and
October 1995 resulting in the identification of 39 additional 
candidates, including the discovery of 9 additional quasars.  
In \S2 we describe the candidate selection process,
in \S3 we describe the spectroscopic observations
and reductions, and in \S4 we discuss the spectroscopic
results.  In \S5, we compare our findings with predictions from 
the determinations of quasar densities from other surveys for quasars.
\S6 contains further discussion.

\section{Candidate Selection}

Candidates were selected from the survey catalog of stellar sources
as described in Paper II.  Briefly, this involves choosing those
objects identified as stellar in a majority of the CCD frames
that comprise the survey and whose colors are consistent with
them being quasars as determined from the colors of
simulated quasar spectra.

In all, 51 previously unidentified candidates were observed.
Twenty-four were bright ultraviolet excess (UVX) objects 
($16.5 < B < 21.0$)
and 13 were fainter UVX objects ($21.0 < B < 22.0$). 
These candidates are expected to have $z<3$.
Candidates for $z>3$ quasars included 5 objects selected as
outliers according to their UBVR magnitudes (VRX) and
nine objects with very red ($B-R$) colors (BRX).

For longslit observations with Cryocam on the KPNO 4m in April 1995,
we used the selection criteria outlined in Paper II to select targets in
the 10h field (which had not been previously investigated spectroscopically),
targets previously observed but with inconclusive spectra,
bright targets requiring only brief exposures,
and most of the remaining VRX and BRX candidates accessible in the
northern spring (most such remaining candidates are fall objects).
Where possible, the slit was rotated and placed so that any other candidates
located within about 2$^{\prime}$ of the primary candidates, and of similar magnitude,
could be observed simultaneously.
Typically these were UVX candidates, as they are the most common.
In fields with no secondary candidate the slit was oriented to observe a
galaxy within 2$^{\prime}$ chosen by C. Liu to calibrate photometric redshifts
calculated for galaxies in these fields (\cite{LIU}).

The primary targets for our October 1995 KPNO 4m run were the BRX and VRX
candidates for which no spectra, or inadequate spectra, existed.
Since most of these are faint and would require long integrations,
we created multislit masks so that as before, secondary candidates
and/or nearby galaxies could be observed simultaneously whenever possible.
Unfortunately the Cryocam CCD was not functioning at this time
and the lower-efficiency RC Spectrograph was used, requiring longer
integration times and preventing us from identifying the faintest candidates.

\section{Spectroscopic Observations and Reductions}
\subsection{April 1995}
We used the Cryocam on the KPNO 4m telescope on UT 1995 April 22-24
to take spectra of observable candidates.
Candidates were chosen for spectroscopy as outlined in \S2.  Conditions
were clear with poor seeing.  

A Loral 800 x 1200 CCD, formatted to 471 x 1200, was used with a gain of
1.5 electrons/ADU.  We chose grism \#770 (300 lines/mm)
and a blocking filter with a blue cutoff at $3850$\AA.  This
gives a dispersion of 4.3\AA/pixel and wavelength coverage from
4500 to 8500\AA.  We used a single slit of 1.7$^{\prime\prime}$
giving a resolution of 15\AA.  
Object exposures ranged from 600s to 2100s, with the most
common exposure time being 1800s.  


The raw two-dimensional CCD frames were corrected for overscan
and bias using the IRAF$^{1}$\footnotetext[1]{The Image
Reduction and Analysis Facility (IRAF) is distributed by the 
National Optical Astronomy Observatories, which is operated by
AURA, Inc., under contract to the National Science Foundation.}
CCDRED package.  Pixel-to-pixel variations
were corrected using a quartz flat field normalized
using the IRAF RESPONSE routine to remove the wavelength dependence
of the flat that is mainly due to the lamp. 
The spectra were then extracted and wavelength and flux calibrated
using the IRAF DOSLIT package with HeNeAr comparison frames and
standard star observations that were obtained each night.

\subsection{October 1995}

As the Cryocam was not available, we used the RC Spectrograph on
UT 1995 October 27-29 to take spectra of candidates chosen from our
fall fields.  Conditions were generally good with some cirrus
on the last night.  Multislit masks were used so that spectra of
secondary candidates could be obtained during the long exposures
needed to detect faint candidates.

A Tektronix 2048$^2$ CCD formatted to 2048 x 651 was used with a gain of 2.0 
electrons/ADU.  We chose grating BL181 (316 lines/mm) and blocking
filter GG455 giving a dispersion of 2.8\AA/pixel and wavelength
coverage of 4100\AA~ranging from 4600\AA~to 10000\AA~depending
on the particular mask and slit position.  The slitmasks had 
2.5$^{\prime\prime}$ slit widths giving a resolution of 10\AA.
Object exposures ranged from 1200s to 2400s.

For each object mask,
typically two object exposures were taken and combined
before extraction.  Quartz flats and HeNeAr
comparison frames were taken between the two object exposures at the
same telescope position.  
Brighter candidates were observed with single shorter exposures in one
slit of a multislit mask or with a single long slit mask, also of 
2.5$^{\prime\prime}$ width.  Quartz flats and
comparisons were obtained for these exposures immediately before
or after the observation at the same telescope position.

The two-dimensional CCD frames were corrected for overscan
and bias using the IRAF CCDRED package.  
For multislit masks or those objects observed through a single aperture
of a multislit mask, the images were corrected for pixel-to-pixel
variations using flat fields created from the quartz lamp images
using the IRAF APFLATTEN routine which removes the wavelength
structure introduced by the quartz lamp.  The spectra were then 
extracted and wavelength calibrated using the TWODSPEC package in 
IRAF.  For those objects observed through the
long slit mask, pixel-to-pixel variations were removed using a 
quartz flat field obtained at the same pointing as the object frame
normalized using the IRAF RESPONSE routine.  The spectra were
then extracted and wavelength calibrated using the IRAF DOSLIT package.
The spectra were not flux calibrated.

\section{Spectroscopic Results}

A total of 57 candidates were observed in April and October, 1995.  Of
these, six were objects previously identified as quasars, but 
with some doubt because the spectra were of low S/N.  All six 
were again identified as quasars.  
Spectroscopy was repeated for nine candidates that were observed 
previously but that could not be identified.  Six of these were
identified from this new spectroscopy, four as galaxies, one as
a CNELG, and one as a star.  
Three others still could not be identified due to the poor S/N level
of the data.  
In total, 39 new candidates were identified:  nine quasars, 4
CNELG's, 8 galaxies, and 18 stars.  Twelve other candidates
were observed but could not be identified.  

\subsection{Quasars}

The list of the nine new quasars found from this spectroscopy is given
in Table 1 including positions, magnitudes, redshifts, and notes.
The objects are numbered following on from Table 1 of Paper II.  
All of these quasars were chosen as UVX candidates.
Their redshifts span the range $0.296 < z < 2.80$ and their
magnitudes $16.8 < B < 21.6$.  This brings the total number of 
quasars found by the DMS survey to date to 55 spanning the
redshift range $0.296 < z < 4.33$.  There are 42 objects with
$z < 2$, nine with $2 < z < 3$, three with $3 < z < 4$, and one
at $z=4.33$.  

Five of the nine quasars reported here have been identified
as quasars on the basis of the presence of one broad emission
line in their spectrum.  In all cases, this line has been
identified as Mg II on the basis of its profile and the absence
of other emission lines.  The other four quasars were 
assigned redshifts on the basis of at least 2 broad emission
lines.  
Of note is DMS 0059$-$0055, the brightest and
lowest redshift quasar found in this survey to date with $z=0.296$ and
$B=16.8$.  Its spectrum, shown in Figure 1, shows prominent 
Fe II emission.  At $z=0.296$ and $B=16.8$, it has $M_B=-23.66$
(for $H_0 = 75~km s^{-1} Mpc^{-1}$, $q_0 = 0.5$, 
and assuming a spectral index of
$\alpha = -0.75$), making it a {\it bona fide} quasar.  
None of the
BRX or VRX candidates observed in 1995 were identified as quasars.

\subsection{Compact Narrow Emission Line Galaxies}

Four of the candidates were identified as compact narrow emission line
galaxies (CNELG's) at redshifts $0.40 < z < 0.66$.  Table 2 lists
their positions, $B$ magnitudes, redshifts, and notes on the objects.
The objects are numbered following on from Table 2 of Paper II.  All of
the objects have weak narrow lines.
As mentioned in Paper II, CNELG's can be a major source of contamination
in UVX surveys for quasars (see Paper II, Figure 9.)  In the initial
spectroscopy reported in Paper II, almost as many CNELG's were recovered
as quasars at $z<3$.  Subsequently, an additional requirement of 
$(B-V) < 0.60$ was adopted in candidate selection to improve the
discrimination between quasars and CNELG's.  The ratio of
CNELG's to UVX quasars recovered from the spectroscopy reported here
is $\sim44\%$, much lower than that reported in Paper II.  Thus this
additional requirement has eliminated a substantial fraction
of these objects, as intended.

\section{Surface Densities}

From the initial spectroscopy, together with the extensive modeling
and simulations presented in Paper II, we estimated that the survey
contained more quasars with $B<21$ and $z<2.3$ than expected from the
KK88 survey and more quasars with $z>3$ than expected
from the results of WHO.
We have recomputed estimated numbers of quasars contained in the DMS
based on our new spectroscopic data (Tables 3A, 3B, and 4). The methods
used are the same as those given in Paper II. Briefly, predicted quasar
counts are computed by multiplying the number of candidate objects we
have selected using a given method by the observational efficiency for
that method (the number of confirmed quasars divided by the number of
spectroscopic IDs).
The expected number of quasars as predicted by previous surveys is
computed by numerically integrating the QLF determined by the 
survey over the redshift and magnitude ranges of interest including 
the detection probabilities of the DMS survey as a function of redshift 
and magnitude as reported in Paper II.

\subsection{z $<$ 3: The UVX Objects}

Observed and expected numbers of UVX quasars in the DMS are given for different
$B$ magnitude bins at $z<2.3$ (Table 3A) and $z>2.3$ (Table 3B.)
The number of candidates in each magnitude bin is listed along with the 
number of spectroscopic identifications, the number of confirmed
quasars, the resulting observational efficiency, and the estimated number
of quasars that are contained in the survey. The 
$1 \sigma$ confidence limits are given using the method 
tabulated by \cite{GEH} for small numbers of
events$^{2}$. \footnotetext[2]{Some $1\sigma$ errors 
presented in Tables 3 and 4 of Paper II were incorrect due to a
small computational error.}
We have recomputed the expected number of quasars using the results
of KK88 and also show the expected number
based on the results of \cite{BSP} (1988, hereafter BSP).

Previously we had computed an observational efficiency of 50\% at $z<2.3$
and $B<21$. Adding our new spectroscopy, this drops to 39.0\% and
decreases the estimated number of quasars from 41.5 to 32.4. 
This result is in good agreement at the $1\sigma$ level with both KK88 and
BSP who predict 29.6 and 32.7, respectively.  Based on the new spectroscopy
and our updated computation of the number of expected quasars from KK88, we see no
evidence for an excess of quasars in our survey at $z<2.3$ and $B<21$.
At $z>2.3$ and $B<21$, our results agree with those of BSP at the $1\sigma$
level and agree with KK88 at the $2\sigma$ level.  We have now identified
over 70\% of the $B<21$ UVX candidates.

Given the decline in the observational efficiencies in this magnitude bin
compared with Paper II, one may wonder if they are likely to decrease
even more as the remaining candidates are observed, thus resulting
in lower predicted numbers of quasars.  However, the remaining candidates do not
occupy regions of color-color space different from those occupied
by the candidates already observed.  Therefore, we do not expect
observational efficiencies to be significantly different for
our remaining candidates.  One issue that was mentioned in 
Paper II and that remains a factor here is that proportionally fewer 
candidates at ($U-B$) $> -0.3$, where quasars at $z>2.3$ are expected, 
have been observed than at ($U-B$) $< -0.3$, where quasars at $z<2.3$ 
are expected (60\% {\it vs.} 80\% for $B < 21.0$.)  
If we compute observational efficiencies 
for these two areas separately, we predict 28.4 quasars at $z<2.3$
(instead of 32.4) and 6.2 quasars at $z>2.3$ (instead of 5.6.)  However,
this is still in good agreement with both KK88 and BSP.  

Estimated and predicted
numbers of UVX quasars in the range $21<B<22$ are in good agreement at
the $1 \sigma$ level at $z>2.3$ and at the $2\sigma$ level at 
$z<2.3$ for both KK88 and BSP and have not changed significantly
since Paper II.  For UVX quasars at $B>22$ we have not completed
enough spectroscopy to give statistically significant results, but
given our average observational efficiency of $\sim35$\% at $z<2.3$, we
are selecting numbers of candidates roughly consistent with the results
of KK88 and BSP.

\subsection{z $>$ 3:  The VRX and BRX Objects}

Table 4 lists observed {\it vs.} predicted numbers of VRX and BRX quasars
at $z>3$. Given is the number of candidates chosen from each method,
the number of candidates for which spectroscopy has been obtained,
the number of confirmed $z>3$ quasars, the observational efficiency,
the resulting estimated number of $z>3$ quasars contained in the
survey, and expected numbers of quasars as predicted by two surveys
for quasars at high redshift. Some $z>3$ quasars were selected both
as VRX and BRX objects.
Of the 12 candidates observed in 1995 but that remain unidentified, 
eight are UVX candidates and four are BRX candidates.  While it is
possible that the unidentified UVX objects are quasars at $z<3$ 
whose redshifts are such that no strong lines are in the wavelength
range covered by the spectroscopy, it is unlikely that the 
unidentified BRX objects are quasars at $z>3$, as the Lyman $\alpha$
line and the drop blueward of the line produce a strong, distinct
feature and should have been detectable in our spectra.
Therefore observational efficiencies were computed by dividing
the number of confirmed quasars by the number of candidates that
have been observed spectroscopically even if the candidate could not
be assigned a firm identification.

The expected numbers of quasars were computed by integrating the QLF
of WHO from $z=3$ to $5$ over the magnitude
range of interest and including DMS detection probabilities.
The last column of Table 4 contains the number of quasars expected from
computing quasar counts as outlined in \cite{KDC} (1995, hereafter KDC).
This involves scaling the \cite{BSP} (1988) $z=2$ QLF down 
in density by
6.5 to match the results of \cite{SSG} (1995, hereafter SSG) 
and KDC at $z=4.4$ and adopting the
evolution predicted by SSG, that is, that quasar space densities fall
off by a factor of 2.7 per unit redshift beyond $z=2.7$.

The WHO survey predicts approximately twice as many quasars as does
KDC for the VRX method. This is because the evolution adopted by KDC is taken from SSG
who predict that quasar space densities begin to decline at $z=2.7$ while
WHO find a peak at $z=3.3$ and a decline beyond.  The DMS VRX detection probabilities
(Paper II, Figure 16) are high at $z=3.0$ to 3.5 where the WHO space densities
are peaking, resulting in higher expected numbers. 
The DMS survey finds one quasar at $V<20.5$ while the WHO and KDC surveys 
predict 2.4 and 1.5 respectively. At $20.5 < V < 22.0$, the DMS survey 
estimates three quasars at $z>3$ while WHO and KDC find 3.8 and 1.9
respectively.  These numbers seem to be in good agreement although the  
numbers involved are small, making any conclusions tentative.

Here, we have combined the four different magnitude bins for the BRX
method from Paper II in to one bin at $17.5 < R < 22.0$.
The DMS has found three confirmed quasars from the 21 observed BRX candidates.
This leads us to predict that the survey contains one additional quasar
in the seven candidates that we have yet to observe for a total of four.
This is down from an estimated number of 8.4 from Paper II.  
Observational efficiencies have dropped from 30\% to 14\%.  However,
the color-color space has now been well explored and the remaining
candidates do not occupy regions of the space different from those
already investigated, so we do not expect this observational efficiency
to change significantly as the remaining candidates are observed.

The number of expected quasars
is higher for KDC than for WHO in the BRX case because while DMS BRX detection
probabilities are rising steeply beyond $z=3.5$ to over 95\% at $z>3.8$, 
WHO predicts that space densities are dropping much more steeply as a function of redshift
than predicted by SSG. 
WHO predict 1.2 quasars while KDC predict 2.0.  
While we still find that the DMS contains an excess of objects at $z>3$
over that expected from the results of WHO and KDC, it is only at the
2 -- 3$\sigma$ level, and, given the small number of objects at $z>3$ contained
in the DMS and that the WHO and KDC results themselves
are based on a relatively small
number of objects, we do not find the excess statistically significant.
On the other hand, it could indicate that 
the decline in quasar space densities at high redshift 
is too steep in both the WHO and SSG cases. Also, we can't rule out the
possibility that there is luminosity-dependent evolution in the QLF
and that the results of both WHO and SSG cannot be extrapolated to the
fainter magnitudes of the DMS. 

\section{Discussion}

Multicolor surveys can suffer from large selection biases that
vary with redshift, magnitude, and spectral energy distribution (SED.)  
We have attempted to model
these biases by producing synthetic quasar spectra with a variety
of properties (see Paper II, \S5
for a detailed description of the synthetic spectra)
and determining the probability of selecting such
quasars given our candidate selection procedures.  The question
then arises:  are the models actually representative of the data?
Figure 2 demonstrates that the theoretical quasar colors computed from the
synthetic spectra do accurately model the colors of the
quasars contained in the DMS.  Shown is a ($B-V$) {\it vs.} ($U-B$)
color-color diagram containing the 27 UVX quasars in the range
($16.5 < B < 21.0$) along with their redshifts. 
Theoretical quasar colors computed
from the synthetic spectra are plotted as tracks.  Four different
quasar SED are shown with varying line strengths and spectral
slopes ($\alpha$) and with redshifts ranging from $z=0.05$ to 3.05
in steps of 0.1.  
The theoretical tracks do bracket the survey 
quasar colors and the trend in color with redshift is also matched
quite well.  In particular, note the two quasars at $z\approx2.8$
at $(B-V)\approx0.3$ and $(U-B)\approx0$.  They would have been
predicted
to be in the redshift range between $z=2.4$ and 2.9 based on their
colors alone.  Since the synthetic quasar spectra play a major role
in computing the survey completeness, the good agreement between 
their colors and those of the real quasars lends
more confidence in our results.  

The discovery of the one bright quasar at $z=0.296$ and $B=16.8$ and the
quasar at $z=4.3$ and $R=20.1$ 
are unexpected in 0.83 deg$^2$, the area of the DMS survey.  We could
expect to find 0.04 quasars at $z<2.2$ and $16.5 < B < 17.0$ in the DMS based
on the QLF of \cite{BSP} (1988), or one every 28 deg$^2$.  
The Large Bright Quasar survey of \cite{HEW} found 15 quasars
with $16.5 < B < 17.0$ in an effective area of 454 deg$^2$, or one every
30 deg$^2$.  The Hamburg/ESO Survey (\cite{HES}) and the
Homogeneous Bright QSO Survey (\cite{LFC}) also predict
one such quasar every $\sim30$ deg$^2$.
Given the agreement of these surveys at these magnitudes, it was indeed
unlikely that this bright object would be found in the DMS.
As for the $z=4.3$ quasar,
we could expect to find 0.06 quasars at $4.0 < z < 4.5$ and $19.5 < R < 20.5$
in the DMS based on the results of WHO or 0.28 based on the results of KDC, or
one quasar every 13 or 3 deg$^2$, respectively.  
While these results are unexpected, due to their small numbers, we do not find
their discovery statistically significant.

The spectroscopy reported here brings the total number of quasars
in the DMS to 55, four of these having $z>3$.  
Given our observational efficiencies including this new spectroscopy, 
we conclude that the number of quasars estimated to be contained in the 
DMS is in good agreement with the results of other surveys for quasars.
The absence of new quasars
with $z > 3$ among the BRX candidates reduces the expected total number
of such quasars in the survey from 8.4 to 4.0.  This
is not a significant excess over the predicted number of 1.2
(allowing for efficiency of detection) based on results by WHO
or of 2.0 based on results by KDC, and, therefore, the estimated number of
objects in the DMS at $z>3$ are in good agreement with these two surveys.
The results for the UVX candidates indicate that the number of quasars in 
the survey with $z < 2.3$ and $B < 22$ is also
less than estimated in Paper II and in good agreement with the
predictions of the KK88 and BSP surveys.

Unfortunately, we were unable to identify any additional UVX candidates
at $B>22.0$ beyond what was reported in Paper II.  This spectroscopy
along with that of the remaining BRX candidates will require an
instrument/telescope combination with greater sensitivity than the
configurations used here.

The next steps are to complete the
spectroscopy of the BRX candidates so that a definitive value of the
surface density of faint $z > 3$ quasars in the survey is established;
complete the spectroscopy of the UVX candidates with $B > 22$ to
constrain the evolution of the faint end of the luminosity function at
$z < 2.3$; make the entire catalog available to the community in
electronic form; and identify and analyze the field galaxies and faint
stars.

\acknowledgements

We would like to thank the KPNO mountain staff for their assistance
in observing and the KPNO TAC for their allocation of time for this
project.  We also thank an anonymous referee for useful comments.
This work was supported in part by NSF Award AST-9529324.

\newpage
\begin{deluxetable}{llcccll}
\tablewidth{38pc}
\small
\tablenum{1}
\tablecaption{Quasars: Identifications}
\tablehead{
\colhead{Number\tablenotemark{a}}        & \colhead{I.D.}        &
\colhead{R.A.(1950)}    & \colhead{Dec.(1950)}  &
\colhead{$B$ mag}         & \colhead{$z$}           &
\colhead{Notes}}
\startdata
47 ......  & DMS 0059$-$0104  & 00{\jh}59{\jm}38{\js}38  &  $-$01{\jd}04{\jam}19{\jas}80  & 
             21.6  & 1.05?  & one line, Mg {\sc ii}? \nl 
48 ......  & DMS 0059$-$0055  & 00{\pjh}59{\pjm}52{\pjs}82  &  $-$00{\pjd}55{\pjam}10{\pjas}73  & 
             16.8  & 0.296  & strong lines \nl 
49 ......  & DMS 1034$-$0040  & 10{\pjh}34{\pjm}33{\pjs}91  &  $-$00{\pjd}40{\pjam}59{\pjas}68  & 
             18.6  & 1.22?  & one line, Mg {\sc ii}? \nl 
50 ......  & DMS 1034$-$0045  & 10{\pjh}34{\pjm}39{\pjs}13  &  $-$00{\pjd}45{\pjam}27{\pjas}26  & 
             19.7  & 0.94?  & one line, Mg {\sc ii}?   \nl 
51 ......  & DMS 1358$-$0101  & 13{\pjh}58{\pjm}18{\pjs}34  &  $-$01{\pjd}01{\pjam}55{\pjas}16  & 
             21.5  & 1.25?  & one line, Mg {\sc ii}?   \nl 
52 ......  & DMS 1358$-$0052  & 13{\pjh}58{\pjm}22{\pjs}70  &  $-$00{\pjd}52{\pjam}18{\pjas}89  & 
             20.4  & 0.85?  & one line, Mg {\sc ii}?  \nl 
53 ......  & DMS 1358$-$0102  & 13{\pjh}58{\pjm}29{\pjs}92  &  $-$01{\pjd}02{\pjam}41{\pjas}64  & 
             20.1  & 0.82   & Mg {\sc ii}, H$\gamma$ \nl 
54 ......  & DMS 1714$+$5012  & 17{\pjh}14{\pjm}21{\pjs}08  &  $+$50{\pjd}12{\pjam}33{\pjas}86  & 
             20.1  & 2.80   & Ly$\alpha$, C {\sc iv} \nl 
55 ......  & DMS 1714$+$5021  & 17{\pjh}14{\pjm}51{\pjs}80  &  $+$50{\pjd}21{\pjam}12{\pjas}67  & 
             21.1  & 1.51   & C {\sc iii}], Mg {\sc ii} \nl 
\enddata
\tablenotetext{a}{The numbers follow on from Table 1 of Paper II}
\end{deluxetable}

\begin{deluxetable}{llcccll}
\small
\tablewidth{38pc}
\tablenum{2}
\tablecaption{Compact Narrow Emission-Line Galaxies: Identifications}
\tablehead{
\colhead{Number\tablenotemark{a}}        & \colhead{I.D.}        &
\colhead{R.A.(1950)}    & \colhead{Dec.(1950)}  &
\colhead{$B$ mag}         & \colhead{$z$}           &
\colhead{Notes}}
\startdata
40 ......  & N0100$-$0101  &  01{\jh}00{\jm}32{\js}11  &  $-$01{\jd}01{\jam}12{\jas}28  &  
             23.9  & 0.67  & ~~~~weak lines  \nl 
41 ......  & N1034$-$0043  &  10{\pjh}34{\pjm}01{\pjs}57  &  $-$00{\pjd}43{\pjam}40{\pjas}07  &  
             21.8  & 0.66  & ~~~~weak lines \nl 
42 ......  & N1358$-$0059  &  13{\pjh}58{\pjm}08{\pjs}90  &  $-$00{\pjd}59{\pjam}48{\pjas}55  &  
             21.9  & 0.400  & ~~~~weak lines \nl 
43 ......  & N2245$-$0214  &  22{\pjh}45{\pjm}46{\pjs}33  &  $-$02{\pjd}14{\pjam}26{\pjas}50  &  
             20.8  & 0.425  & ~~~~weak lines \nl 
\enddata
\tablenotetext{a}{The numbers follow on from Table 2 of Paper II}
\end{deluxetable}

\begin{deluxetable}{cccccccc}
\tablenum{3a}
\footnotesize
\tablecaption{Observed versus Predicted Numbers of $z<2.3$ UVX Quasars}
\tablehead{
\colhead{} &
\colhead{} &
\colhead{} &
\colhead{} &
\colhead{} &
\colhead{Estimated} &
\multicolumn{2}{c}{Expected No.} \nl
\colhead{} &
\colhead{} &
\colhead{} &
\colhead{Confirmed} &
\colhead{} &
\colhead{Number} &
\multicolumn{2}{c}{of $z<2.3$ Quasars\tablenotemark{a}} \nl
\cline{7-8}
\colhead{Magnitude Bin} &
\colhead{No. of UVX} &
\colhead{No. of } &
\colhead{$z<2.3$} &
\colhead{Observational} &
\colhead{of $z<2.3$} &
\multicolumn{2}{c}{} \nl
\colhead{Limits} &
\colhead{Candidates} &
\colhead{UVX ID's} &
\colhead{Quasars} &
\colhead{Efficiency} &
\colhead{Quasars} &
\colhead{KK88} &
\colhead{BSP} }
\startdata
16.5 $< B <$ 21.0 & \phn83 &    59 & 23 & $39.0^{+\phn7.3}_{-\phn6.9}\%$ & $32.4^{+\phn3.0}_{-\phn2.9}$ & 29.6 & 32.7  \nl \nl
21.0 $< B <$ 22.0 & \phn71 &    33 & 17 & $51.5^{+10.0}_{-10.1}$\ppc & $36.6\pm 4.9$ &  41.8 & 31.0 \nl \nl
22.0 $< B <$ 22.6 & \phn99 &    10 & \phn2 &  $20.0^{+20.7}_{-12.8}$\ppc  & $19.8^{+18.8}_{-12.0}$ &  48.2 & 26.0\nl \nl
22.6 $< B <$ 23.1 &    145 & \phn0 & \phn0 &  ... & ... & 42.8 & 18.4 \nl 
\enddata
\tablenotetext{a}{Calculated by convolving the number-magnitude relation
of Koo \& Kron 1988 (KK88) and the luminosity function of 
Boyle, Shanks, \& Peterson 1988 (BSP) with the UVX detection 
probabilities as discussed in the text.}
\end{deluxetable}

\begin{deluxetable}{cccccccc}
\tablenum{3b}
\footnotesize
\tablecaption{Observed versus Predicted Numbers of $z>2.3$ UVX Quasars}
\tablehead{
\colhead{} &
\colhead{} &
\colhead{} &
\colhead{} &
\colhead{} &
\colhead{Estimated} &
\multicolumn{2}{c}{Expected No.} \nl
\colhead{} &
\colhead{} &
\colhead{} &
\colhead{Confirmed} &
\colhead{} &
\colhead{Number} &
\multicolumn{2}{c}{of $z>2.3$ Quasars\tablenotemark{a}} \nl
\cline{7-8}
\colhead{Magnitude Bin} &
\colhead{No. of UVX} &
\colhead{No. of } &
\colhead{$z>2.3$} &
\colhead{Observational} &
\colhead{of $z>2.3$} &
\multicolumn{2}{c}{} \nl
\colhead{Limits} &
\colhead{Candidates} &
\colhead{UVX ID's} &
\colhead{Quasars} &
\colhead{Efficiency} &
\colhead{Quasars} &
\colhead{KK88} &
\colhead{BPS} }
\startdata
16.5 $< B <$ 21.0 & \phn83 &    59 & 4 & $6.8^{+\phn5.1}_{-\phn3.2}\%$ & $5.6^{+\phn1.7}_{-\phn1.5}$ & 3.4 & 4.4  \nl \nl
21.0 $< B <$ 22.0 & \phn71 &    33 & 1 & $3.0^{+\phn6.9}_{-\phn2.4}$\ppc & $2.2^{+\phn2.8}_{-\phn1.4}$ &  3.6 & 3.8 \nl \nl
22.0 $< B <$ 22.6 & \phn99 &    10 & 0 & $0.0^{+16.8}_{-\phn0.0}$\ppc  & $0.0^{+15.0}_{-\phn0.0}$ &  3.3 & 2.3\nl \nl
22.6 $< B <$ 23.1 &    145 & \phn0 & 0 &  ... & ... & 1.4 & 0.9 \nl 
\enddata
\tablenotetext{a}{Calculated by convolving the number-magnitude relation
of Koo \& Kron 1988 (KK88) and the luminosity function of 
Boyle, Shanks, \& Peterson 1988 (BSP) with the UVX detection 
probabilities as discussed in the text.}

\end{deluxetable}

\begin{deluxetable}{ccccccccc}
\tablewidth{43pc}
\small
\tablenum{4}
\footnotesize
\tablecaption{Observed versus Predicted Numbers of VRX and BRX Quasars}
\tablehead{
\colhead{} & \colhead{} & \colhead{} & \colhead{} & \colhead{} & \colhead{} & \colhead{Estimated} & 
\multicolumn{2}{c}{Expected Number} \nl
\colhead{} & \colhead{} & \colhead{} & \colhead{} & \colhead{Confirmed} & \colhead{} & \colhead{Number} & 
\multicolumn{2}{c}{of $z>3$ Quasars} \nl
\cline{8-9}
\colhead{Selection} & \colhead{Magnitude} & \colhead{No. of} & \colhead{No.
of} & \colhead{$z>3$} & \colhead{Observational} & \colhead{of $z>3$} & 
\multicolumn{2}{c}{} \nl
\colhead{Method} & \colhead{Bin Limits} & \colhead{Candidates} & \colhead{Spectra} & 
\colhead{Quasars} & \colhead{Efficiency} & \colhead{Quasars} & \colhead{WHO\tablenotemark{c}} & \colhead{KDC\tablenotemark{d}}}
\startdata
VRX\tablenotemark{a}~~~~ & $17.5 < V < 20.5$ &  \phn3 & \phn3 & 1 & 
$33.3^{+41.5}_{-26.7}\%$ & $1.0\pm0.0$ & 2.4 & 1.5 \nl\nl
VRX\tablenotemark{a}~~~~ & $20.5 < V < 22.0$ &  \phn3 & \phn2 & 2 & 
$100\pm0.0$\ppc  & $3.0^{+0.0}_{-1.0}$ & 3.8 & 1.9 \nl\nl
BRX\tablenotemark{b}~~~~ & $17.5 < R < 22.0$ &  28 &       21 & 3 & 
$14.3^{+12.1}_{-\phn7.7}$\ppc & $4.0^{+1.3}_{-1.1}$ & 1.2 & 2.0 \nl
\enddata
\tablenotetext{a}{Based on $(U-V)/(V-R) + (B-V)/(V-R)$ color-color
diagrams}
\tablenotetext{b}{Based on $(B-R)/(R-I86) + (B-R)/(I75-I86)$ color-color
diagrams}
\tablenotetext{c}{Calculated using the luminosity function of 
Warren, Hewett, \& Osmer 1994 (WHO) and the survey detection
probabilities.}
\tablenotetext{d}{Calculated using the survey detection
probabilities and the luminosity function of
Boyle, Shanks, and Peterson 1988 scaled to the results of 
Kennefick {\it et al.} 1995 (KDC) and Schmidt, Schneider, 
\& Gunn 1995 (SSG) at $z\sim4$ and adopting the SSG form
of evolution in quasar space densities.}
\end{deluxetable}

\begin{figure}
\plotone{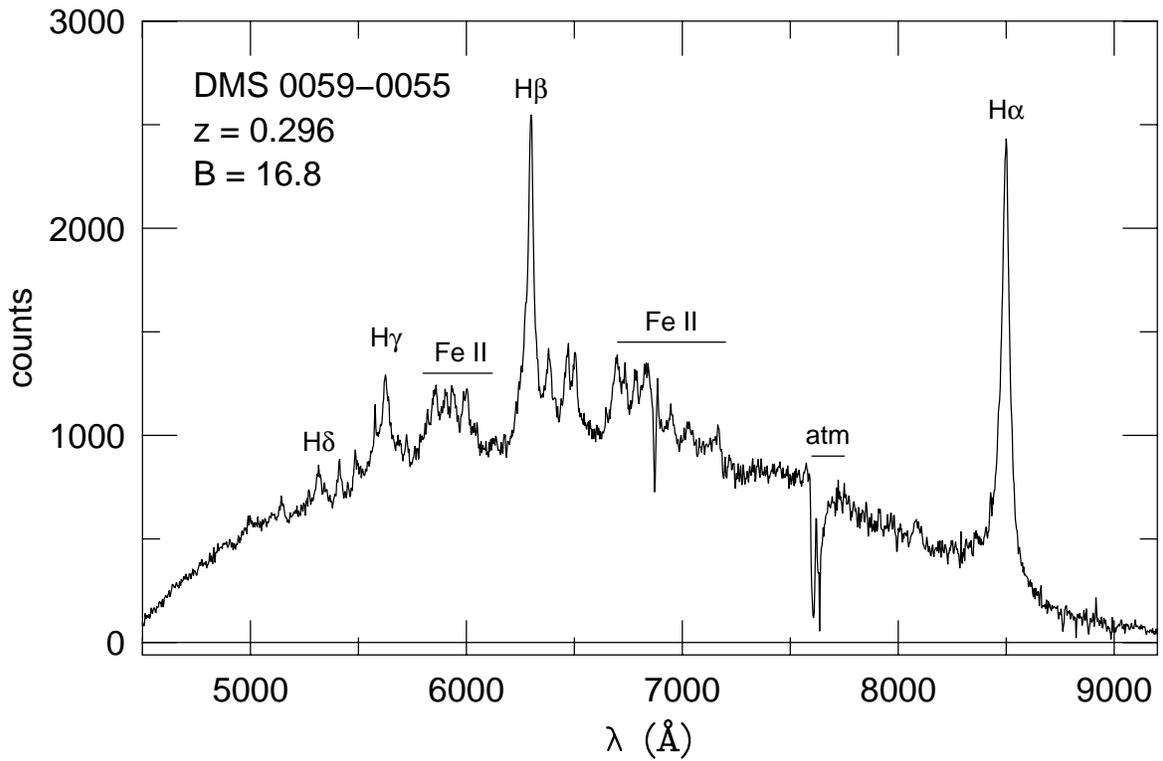}
\caption{A spectrum of DMS 0059$-$0055, a bright 
quasar at $z=0.296$.
The 900 second spectrum was obtained with the RC Spectrograph at the
KPNO 4 meter telescope on UT 1995 October 28.  This is the brightest
and lowest redshift quasar in the survey.}
\end{figure}

\begin{figure}
\plotone{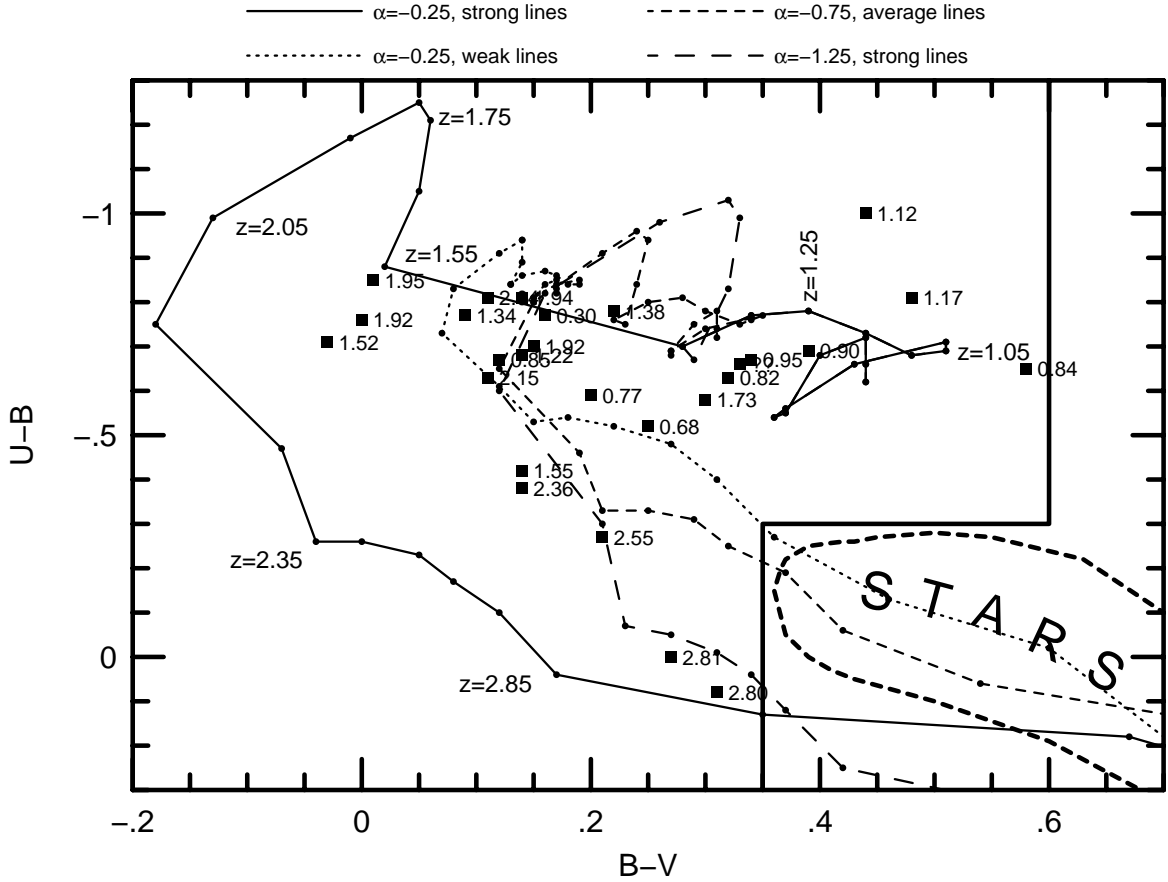}
\caption{A ($B-V$) {\it vs.} ($U-B$) color-color 
diagram showing the 
27 UVX survey quasars in the range ($16.5 < B < 21.0$) (filled squares).
The redshift of the quasar is given to the right of the symbol.  Also 
shown are the candidate selection limits (heavy solid line) and the
approximate location of the stellar locus.  Theoretical color-redshift 
tracks are also shown for a variety of quasar continuum slopes ($\alpha$)
and emission line strengths as indicated above the
diagram.  Average line strengths were taken from \cite{WIL}
with weak lines taken as half the equivalent width of average
lines and strong lines taken as twice the equivalent width.
The tracks as shown cover the redshift range from 0.05 to 3.05
in steps of 0.1 (dots).  The redshift of a few points is indicated
for one track (solid line, $\alpha=-0.25$, strong emission lines) and
should allow the reader to infer the redshifts for points on the
other tracks based on the shape of the tracks. 
For further discussion, see \S6.}
\end{figure}

\end{document}